\documentclass[aps,notitlepage,11pt]{revtex4-1}
\usepackage{amssymb}
\usepackage{natbib}
\usepackage{amsmath}
\usepackage{amsfonts}
\usepackage{graphicx}
\usepackage{mathrsfs}
\usepackage{dcolumn}
\usepackage{bm}
\usepackage{color}
\usepackage{cancel}
\usepackage{mathbbol}
\usepackage{epsfig}
\usepackage{units}
\usepackage{esint}

\usepackage{%
  graphicx,
  textcomp,
  dcolumn,% Align table columns on decimal point
  natbib,%
  amsmath
}

\usepackage[range-units= single,range-phrase= -]{siunitx}

\newcommand{\eq}[2]{\begin{equation}\label{#1}#2\end{equation}}

\newcommand{\abs}[1]{\left|#1\right|}
\newcommand{\rk}[1]{\left( #1\right)}

\newcommand{\rem}[1]{}
\newcommand{\tn}[1]{\textnormal{#1}}

\newcommand{\ket}[1]{\left| #1 \right>}
\begin{document}

\title{Phase dependent loading of Bloch bands and Quantum simulation of relativistic wave equation predictions with ultracold atoms in variably shaped optical lattice potentials}

\author{Christopher Grossert}
\email{grossert@iap.uni-bonn.de}
\author{Martin Leder}
\author{Martin Weitz}

\affiliation{Institut f\"ur Angewandte Physik der Universit\"at Bonn, Wegelerstr. 8, 53115 Bonn}
\date{\today}

\begin{abstract}
The dispersion relation of ultracold atoms in variably shaped optical lattices can be tuned to resemble that of a relativistic particle, i.e. be linear instead of the usual nonrelativistic quadratic dispersion relation of a free atom. Cold atoms in such a lattice can be used to carry out quantum simulations of relativistic wave equation predictions. We begin this article by describing a Raman technique that allows to selectively load atoms into a desired Bloch band of the lattice near a band crossing. Subsequently, we review two recent experiments with quasirelativistic rubidium atoms in a bichromatic lattice, demonstrating the analogs of Klein tunneling and Veselago lensing with ultracold atoms respectively.
\end{abstract}

\maketitle

%%%%%%%%%%%%%%%%%%%%%%%%%%%%%%%%%%%%%%%%%%%%%
\section{Introduction}

The Dirac equation, as a relativistic physics quantum wave equation, was developed very soon after the introduction of nonrelativistic quantum mechanics \cite{dirac}. This equation correctly predicts the fine structure corrections of atoms and is a starting point of modern quantum field theory. It is interesting that basic predictions of the Dirac equation, as Klein tunneling \cite{klein1,bjorken} or Zitterbewegung \cite{zitterbewegung1}, have never been observed for elementary particles. For Klein tunneling, an effect frequently discussed in textbooks where relativistic particles penetrate through a potential barrier without the exponential damping that occurs for relativistic quantum tunneling, for electrons an electric field strength of order $10^{16}$V/cm would be required for an observation, an issue that so far has prevented experimental realization for these particles. On the other hand, quantum simulations of Klein tunneling, as well as other relativistic wave equation predictions as Zitterbewegung, are well feasable \cite{graphen2,iontrap}. The background is that with suitable low energy systems the corresponding relativistic Hamiltonians can be engineered, with the dispersion being linear (or near linear) in a certain energy range and the effective speed of light being orders of magnitude below the true speed of light in vacuum. Indeed, also the simulation of high energy physics quantum field theories has been proposed with ultracold atoms of temperatures in the nano-Kelvin range \cite{QFT}.

Experimentally, the analog of Klein tunneling has been observed in solid state physics systems, as graphene \cite{graphen1,graphen2}. Other systems for the observation of quasirelativistic effects are ions in Paul traps \cite{iontrap,gerritsma}, see also work relating to photonic structures \cite{beenakker,zhang} and dark state media \cite{darkstate}. Here we are interested in ultracold atoms for quantum simulation of relativistic physics predictions. We review an experiment demonstrating the analog of Klein tunneling with ultracold atoms \cite{salger11}. More recently, Dirac points have been generated in a two-dimensional lattice with fermionic atoms \cite{esslinger} and the analog of Zitterbewegung has been observed \cite{zitterbewegung2}. We also review a more recent demonstration of Veselago lensing for matter waves carried out in our group \cite{leder14}, an effect where a spatially diverging pencil of rays is focused to a spatially converging one using negative refraction. The experimental realization is also performed using ultracold atoms in the lattice. Further, we describe a method to selectively load atoms into certain Bloch bands of an optical lattice near an avoided crossing using a sequence of two simultaneously performed Bragg pulses with well defined relative phase.

The paper is organized as follows. We begin in section 2 by introducing the general background of our experiments relying on cold rubidium atoms in a bichromatic optical lattice. We derive an effective Hamiltonian for the cold atoms in the lattice that resembles a one-dimensional Dirac Hamiltonian. We next describe a method to phase selectively load into Bloch bands. Section 4 then describes experimental results demonstrating the analog of Klein tunneling with cold atoms in the lattice, while section 5 gives results on the realisation of an 1D Veselago lens. Finally, section 6 gives conclusions and an outlook. 

\section{Experimental background}

\subsection{Experimental sequence}
Our experiment is based on ultracold rubidium atoms in a Fourier synthesized optical lattice potential, which allows to tailor the dispersion relation. Initially, a Bose-Einstein condensate of rubidium atoms ($^{87}$Rb) is created by evaporative cooling of atoms in a far detuned optical dipole trap. During the final stage of evaporative cooling, an additional magnetic field gradient is used to selectively prepare the atoms in the $\ket{F=1,\ m_F=0}$ Zeeman state. Subsequently the dipole trapping potential is switched off. Next, the optical lattice potential (and in some experiments also additional, spatially slowly varying dipole potentials) are switched on and the system evolves with the respective dynamics. At the end of such an experimental cycle, all dipole potentials are switched off and the atoms expand freely in the earth's gravitational field. The spatial atomic distribution can be measured by taking an absorption image either immediately afterwards (position analysis) or alternatively after a free expansion time of typically $\SIrange{10}{20}{ms}$ (momentum analysis). The Zeeman state of atoms can be analysed by either applying a Stern-Gerlach force with a magnetic field gradient, which allows one to spatially resolve atoms in the different Zeeman states in a far field time-of-flight measurement, or by using a Zeeman state-selective microwave $\pi$-pulse which transfers the atoms from the $\ket{F=1,\ m_F}$ to the $\ket{F=2,\ m_F'}$ hyperfine state and subsequently recording a shadow image on a CCD-camera with a laser beam pulse tuned to the $F=2\rightarrow F'=3$ hyperfine component of the Rubidium D2-line.

\subsection{Fourier synthesized optical lattices}
\begin{figure}
  \centering
  \includegraphics[width=0.9\textwidth]{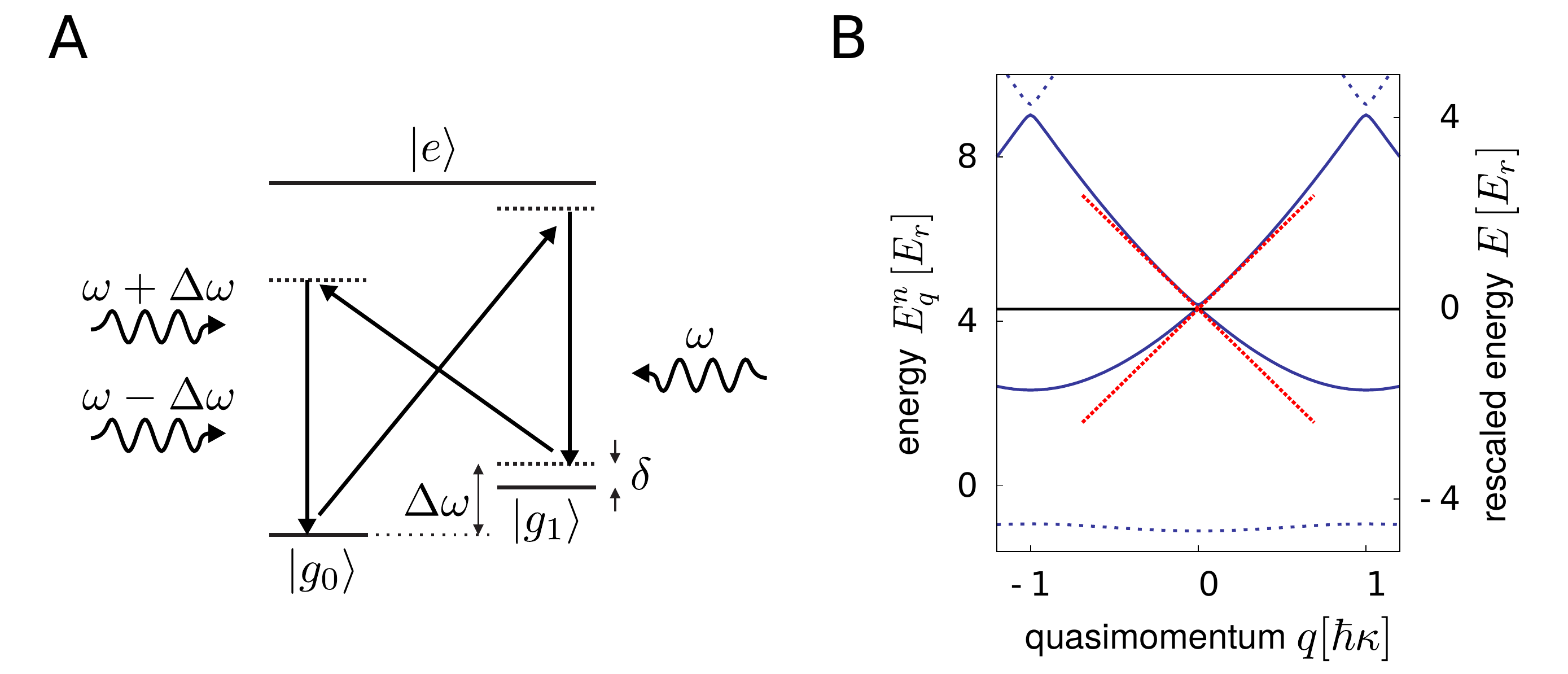}
  \caption{\textbf{A}: Four-photon scheme for generating an optical lattice potential with $\lambda/4$ spatial periodicity based on a three-level atom driven by an optical field of frequency $\omega$ and two counterpropagating optical fields of frequencies $\omega\pm \Delta \omega$. \textbf{B}: Energy dispersion relation for the combined optical lattice potential obtained by superimposing lattice potentials of $\lambda/2$ and $\lambda/4$ spatial periodicities for a relative phase of $\pi$ between lattice harmonics and $V_1=4E_r$, $V_2=1E_r$. The first and second excited Bloch band (blue solid lines) touch each other at $q=0\hbar \kappa$. Near this crossing, the dispersion relation is linear, as indicated by the red lines. On the right hand side, the energy measured relatively to that of the crossing is shown.}
  \label{fig:1}
\end{figure}

A one-dimensional biharmonic lattice potential is imprinted to the ultracold rubidium atoms \cite{Ritt}. For the fundamental spatial frequency of the variable lattice, we use a usual standing wave lattice of spatial periodicity $\lambda/2$, generated by two counterpropagating beams of same optical frequency $\omega$. Here $\lambda$ denotes the wavelength of the driving laser beam. Higher order harmonics are accessable by multiphoton Raman processes \cite{Ritt}. We here make use of a four-photon Raman process, as realized with a set of two copropagating beams of frequencies $\omega+\Delta\omega$ and $\omega-\Delta\omega$ respectively and a counterpropagating beam of frequency $\omega$ in a three-level scheme with two stable ground states $\ket{g_0}$ and $\ket{g_1}$ and an electronically excited manifold $\ket{e}$, see Fig. \ref{fig:1}A. The dispersion of the induced four-photon processes generates a lattice potential of spatial periodicity $\lambda/4$. By superimposing lattice potentials with spatial periodicities $\lambda/2$ and $\lambda/4$ respectively with a suitable relative phase between lattice harmonics, both spatially symmetric and asymmetric Fourier synthesized lattice potentials can be generated. All required optical frequencies to synthesize the variable lattice are derived from a single laser source with wavelength detuned $\SI{3}{nm}$ to the red of the rubidium D2-line.

\subsection{Band structure and relativistic energy dispersion}

The dynamics of ultracold atoms in a bichromatic optical lattice along the $z$-direction is described by the Hamiltonian
\eq{eq:H1}{H_0 \rk{z}=\frac{-\hbar^2\partial_z^2}{2m}+V_1/2\cos\rk{2\kappa z}+V_2/2\cos\rk{4\kappa z+\phi},}
where $\kappa=2\pi/\lambda$ corresponds to the optical wave vector, $m$ is the rubidium atomic mass, $V_{1,2}$ denote the potential depths and $\phi$ the relative phase between the harmonics of the lattice potentials with spatial periodicities $\lambda/2$ and $\lambda/4$, respectively. In addition, we allow for a spatially slowly varying potential $V_z\rk{z}$ as can be e.g. be imprinted by an additional dipole trapping beam, a magnetic field gradient or in the atomic rest frame by a slow acceleration of the optical lattice. Solving the stationary Schr\"odinger equation
\eq{eq:sgl}{H_0\rk{z}\psi\rk{z}=E_q^n\psi\rk{z}}
is done using a Bloch ansatz, which reduces the problem to the eigenvalue equation
\eq{eq:H2}{\sum_l H_{j,l}b_l=E_q^nb_l,}
with $H_{i,j}$ denoting the corresponding Hamiltonian in matrix form. When we restrict ourselves to the ground and the first two excited Bloch bands, the Hamiltonian reads
\eq{eq:H3}{H =
\rk{\begin{array}{ccc}
  \rk{q/\hbar \kappa-2}^2E_r  & V_1/4 & V_2/4\exp\rk{i\phi} \\
  V_1/4 & \rk{q/\hbar \kappa}^2 E_r & V_1/4  \\
   V_2/4\exp\rk{-i\phi} & V_1/4 & \rk{q/\hbar \kappa+2}^2E_r 
\end{array}},
}
where $E_r=(\hbar\kappa)^2/2m$ denotes the atomic recoil energy. As shown in \cite{salger11}, in vicinity of the avoided crossing between the first and second excited Bloch bands, the ground band can be adiabatically eliminated. After a basis transformation, the effective Hamiltonian can be written as 
\eq{eq:Heff}{H_{\tn{eff}} = m_{\tn{eff}}c_{\tn{eff}}^2\sigma_z+c_{\tn{eff}}q\sigma_x,}
with the Pauli matrices $\sigma_{i}$, an effective speed of light $c_{\tn{eff}}=2\hbar \kappa/m\approx \SI{11}{mm/s}$, and an effective mass which is given as
\eq{eq:meff}{m_{\tn{eff}}=\frac{\Delta E}{2c_{\tn{eff}}^2}=\frac 1 {c_{\tn{eff}}^2}\abs{\frac {V_1}{64E_r}+\frac{V_2}{4}e^{i\phi}}.}
Here, $\Delta E$ denotes the energy splitting between the first two excited Bloch bands, which is determined by the lattice parameters $V_{1,2}$ and $\phi$. For $\Delta E=0$, the effective Hamiltonian $H_{\tn{eff}}$ of eq. (\ref{eq:Heff}) shows that the dispersion near the crossing is linear, see also Fig. \ref{fig:1}B. For both vanishing and small splittings between the Bloch bands, the cold atom system enables us to study relativistic physics effects predicted by the $1+1$ Dirac equation, though the true speed of the particles is many orders of magnitude below the speed of light in vacuum.

\section{Phase-dependent diabatic loading into Bloch states}

\begin{figure}
  \centering
  \includegraphics[width=0.9\textwidth]{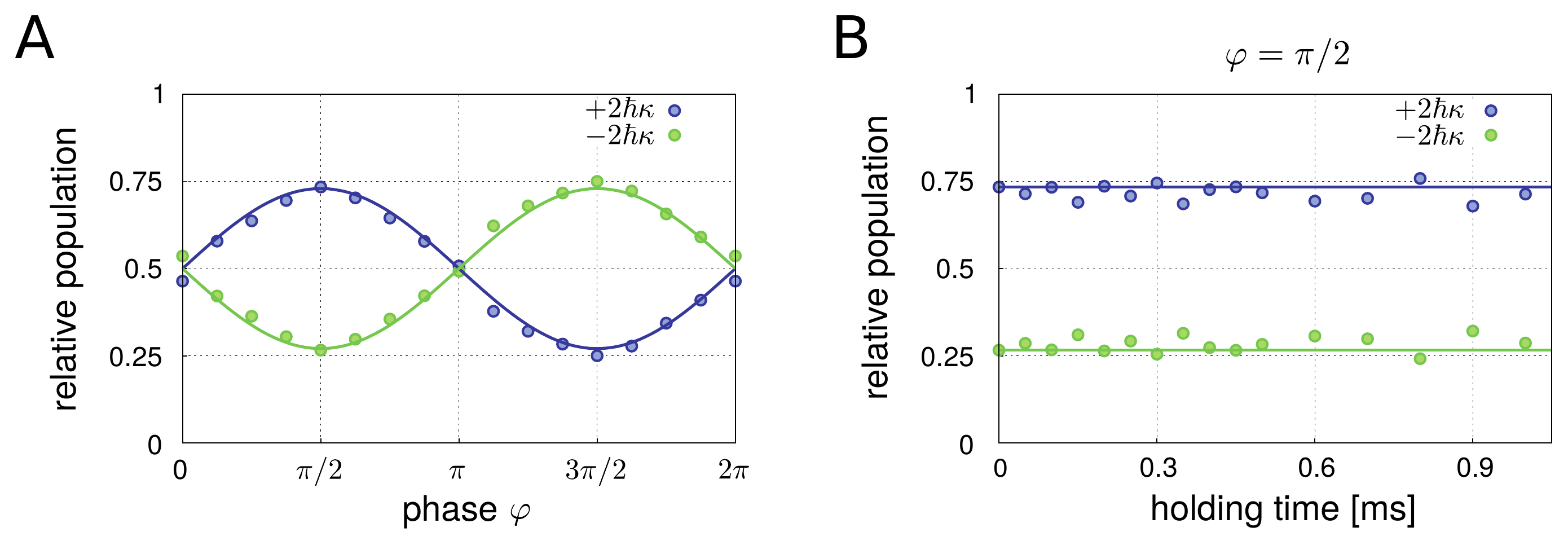}
  \caption{Diabatic loading into Bloch states. The shown diagrams give the observed relative atomic population in the $\ket{+2\hbar \kappa}$ and $\ket{-2\hbar \kappa}$ momentum states after adiabatically accelerating away from the band crossing, as a measure for the population in upper and lower Bloch bands respectively. \textbf{A}: Dependence of the observed bands population on the relative phase between Bragg pulses. The shown lines are a guide for the eyes. \textbf{B}: Dependence of band population on the length of an additional holding time for a relative phase $\varphi=\pi/2$ between Bragg pulses.}
  \label{fig:2}
\end{figure}

We next describe experiments investigating the selective loading of atoms into a desired Bloch state near a band crossing. We restrict ourself to the first two excited Bloch bands of the rubidium atomic lattice system, assuming that the lattice is not too deep such that coupling to other bands near a crossing of these two bands is small. The eigenstates of these two bands can now be expressed in terms of the eigenstates $\ket{-2\hbar \kappa}$ and $\ket{+2\hbar \kappa}$ of the free atom. When diagonalizing the effective Hamiltonian (\ref{eq:Heff}) assuming a finite size energy gap $\Delta E \not = 0$ one readily finds

\begin{align}
  \ket{+} &= +\cos\rk{\theta/2} \ket{+2\hbar \kappa} +\sin\rk{\theta/2} \ket{-2\hbar \kappa} \label{eq:pm1},\\
  \ket{-} &= -\sin\rk{\theta/2} \ket{+2\hbar \kappa} +\cos\rk{\theta/2} \ket{-2\hbar \kappa} \label{eq:pm2},
  \intertext{where the mixing angle $\theta$ is given by}
  \tan\rk{\theta} &= \frac{\abs{m_{\tn{eff}}c_{\tn{eff}}}}{q}, \quad 0\le \theta \le \pi \label{eq:tan}. 
\end{align}
One way to selectively load atoms into the upper ($\ket{+}$) or lower ($\ket{-}$) eigenstate is to initially load the atoms into one band far from the crossing, where the mixing angle between the two bands is small and selective loading can be achieved since a free atomic eigenstate of certain momentum has almost full overlap with an eigenstate in the lattice (again, we assume a not too deep lattice). Subsequently, the atomic population can be transferred adiabatically, e.g. by corresponding slow acceleration of the lattice, into the region of the crossing (see e.g. \cite{salger11}).

In this section, we describe a method to diabatically load into the eigenstates of the lattice by transferring atoms initially in the ground band to a desired upper band near a crossing by simultaneously imparting two Bragg pulses to the atoms, each of which resonantly couple ground band atoms (in state $\ket{0}$) to the eigenstates $\ket{-2\hbar \kappa}$ and $\ket{+2\hbar \kappa}$ respectively. For equal pulse areas of both pulses and a relative phase $\varphi$ of the two Bragg pulses, the part of the atomic wavefunction left in the two upper Bloch bands is in the coherent superposition

\eq{eq:blochrep}{\ket{\psi}=\frac{1}{\sqrt{2}}\rk{\ket{2 \hbar \kappa}+e^{i\varphi}\ket{-2 \hbar \kappa}}.}

The overlap of $\ket{\psi}$ to the eigenstates $\ket{+}$ and $\ket{-}$ (\ref{eq:pm1},\ref{eq:pm2}) now determines the loading efficiency into the two corresponding Bloch bands, and a suitable choice of the relative phase $\varphi$ of the two Bragg pulses can be used as a mean to control the relative population of the lattice eigenstates $\ket{+}$ and $\ket{-}$. To determine which of the two eigenstates $\ket{\pm}$ is populated, we perform an adiabatic acceleration to $0.5\hbar k$ away from the energy band crossing and then determine the atomic momentum distribution by a time-of-flight technique. For atoms in the upper excited Bloch band (asuming a small splitting) this will then lead to a population of the $\ket{+2\hbar k}$ momentum state, while atoms in the lower excited Bloch band will be rotated to $\ket{-2\hbar k}$.

Fig. \ref{fig:2}A shows experimental results for the dependency of the relative eigenstate population on the phase $\varphi$ between the two Bragg pulses. This experiment was carried out only with a four-photon lattice present, i.e. without the presence of a standing wave lattice potential, and the lattice depth of the four-photon lattice was $\approx 1.0\ E_r$. The observed relative population of the $\ket{\pm 2\hbar \kappa}$ states shows a sinusoidal dependence on the phase $\varphi$, as expected. The observed finite contrast of the sinusoidal variation is attributed to remaining coupling to other bands, the finite initial momentum width of the atomic sample, and the limited Rabi frequencies of the Bragg pulses.

In a further experiment, we have included an additional, variable holding time before accelerating away from the band crossing and performing time-of-flight imaging for detection to verify whether the two superimposed Bragg pulses leave the atoms in an eigenstate of the system. Fig. \ref{fig:2}B shows the observed dependence of the relative population on the holding time. For the shown parameter range below $\SI{1}{ms}$, the relative population of the Bloch states remains constant, as expected for an eigenstate. For longer holding times, we observe atom loss attributed to spontaneous scattering from the lattice beams, which prohibits further detailed investigations.

\section{Klein tunneling}

Non-relativistic quantum tunneling corresponds to the effect that a particle is able to penetrate a potential barrier which is not surmountable in the classical case with an exponential loss of the probability amplitude. For relativistic particles, on the other hand, also penetration of a barrier without this exponential damping is possible. This effect, as predicted by Klein \cite{klein1}, occurs when a strong potential, being repulsive for particles and attractive for antiparticles, results in particle and antiparticle states matching in energy across the barrier. This allows for a high transmission probability, when a potential drop of the order of the particle rest energy $mc^2$ is achieved over the Compton length $h/mc$. Here we describe an experiment observing the analog of Klein tunneling for cold atoms \cite{salger11}, where an optical lattice is used to imprint a relativistic dispersion relation. In other words, we perform a quantum simulation of a relativistic wave equation prediction. As in other low energy realizations, as in solid state analogons as graphene \cite{graphen1,graphen2} or in ion traps \cite{iontrap}, the basic idea here is that the linear dispersion corresponds to an effective speed of light many orders of magnitude below $c$, which decreases the required potential drop.

\begin{figure}
  \centering
  \includegraphics[width=0.9\textwidth]{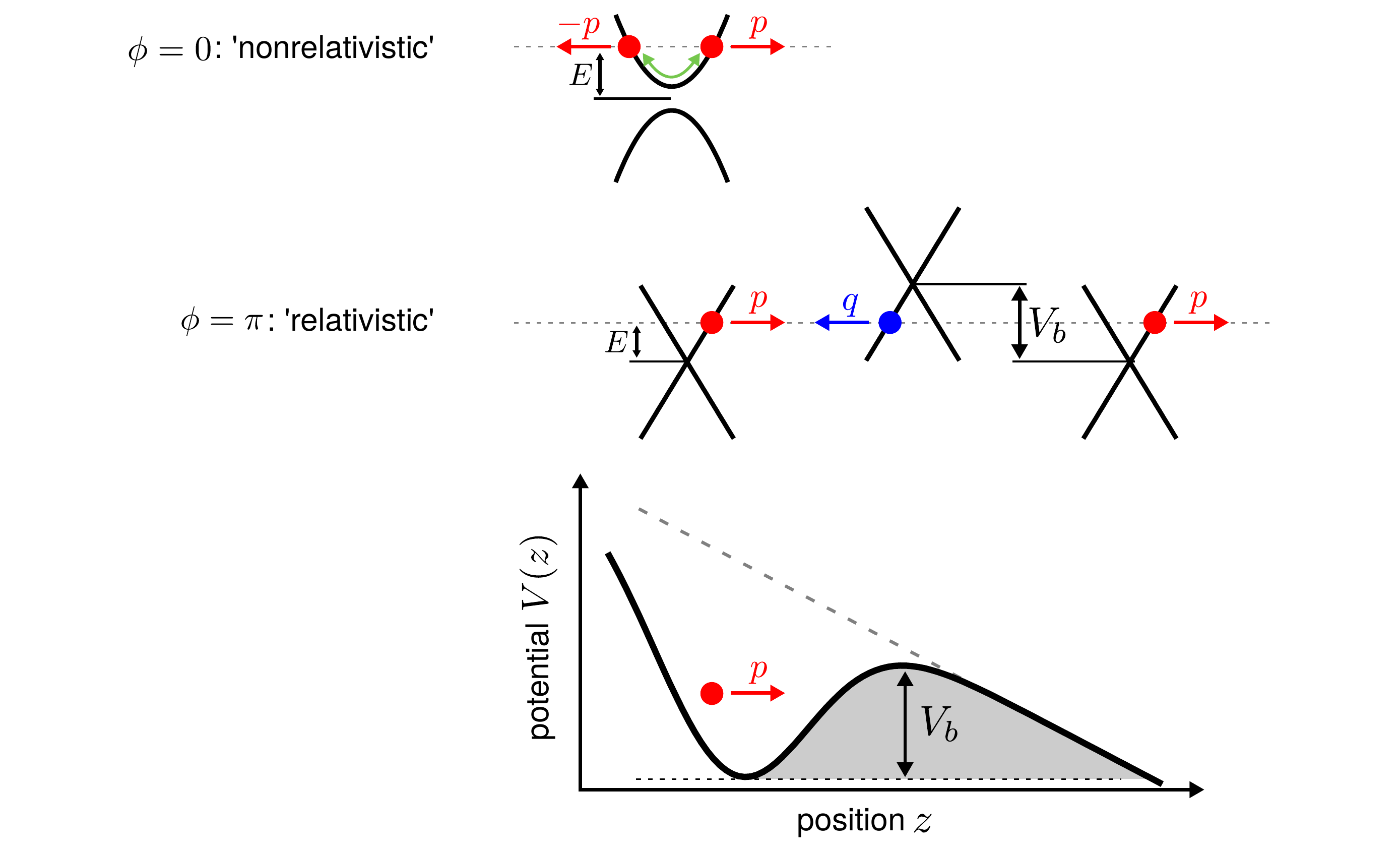}
  \caption{Experimental scheme for investigating Klein tunneling. Atoms with an initial momentum $p$ approach a potential barrier of height $V_b$. The top plot shows the relevant part of the dispersion relation in the lattice for a relative phase between lattice harmonics of $\phi=0$, for which the large splitting between bands suppresses a tunneling between bands. In contrast, for $\phi=\pi$ (middle plot) the dispersion is linear and the bands touch. Atoms can then overcome the barrier due to their possibility to drop below the band crossing in the Bloch spectrum when loosing potential energy.}
  \label{fig:3}
\end{figure}

For the dispersion imprinted by our bichromatic optical lattice potential for cold atoms, see section 2. The effective speed of light that is relevant to the experiments is $c_{\tn{eff}}\approx \SI{1,1}{cm/s}\approx 10^{-10}c$. We can both investigate the case of an effectively relativistic dispersion near the position of the crossing between the first and the second excited Bloch band, for the case of a vanishing or small splitting between bands, or the case of a quadratic dispersion with atoms not being able to tunnel between bands for the case of a large splitting between bands, (see also Fig. \ref{fig:3} middle and top plot respectively). Klein tunneling of atoms is investigated by monitoring the transmission of cold atoms through a barrier of width and height too large to allow for usual quantum tunneling, the barrier being realized by the combined action of a far detuned optical dipole trap potential and the gravitational potential.  

\begin{figure}
  \centering
  \includegraphics[width=0.9\textwidth]{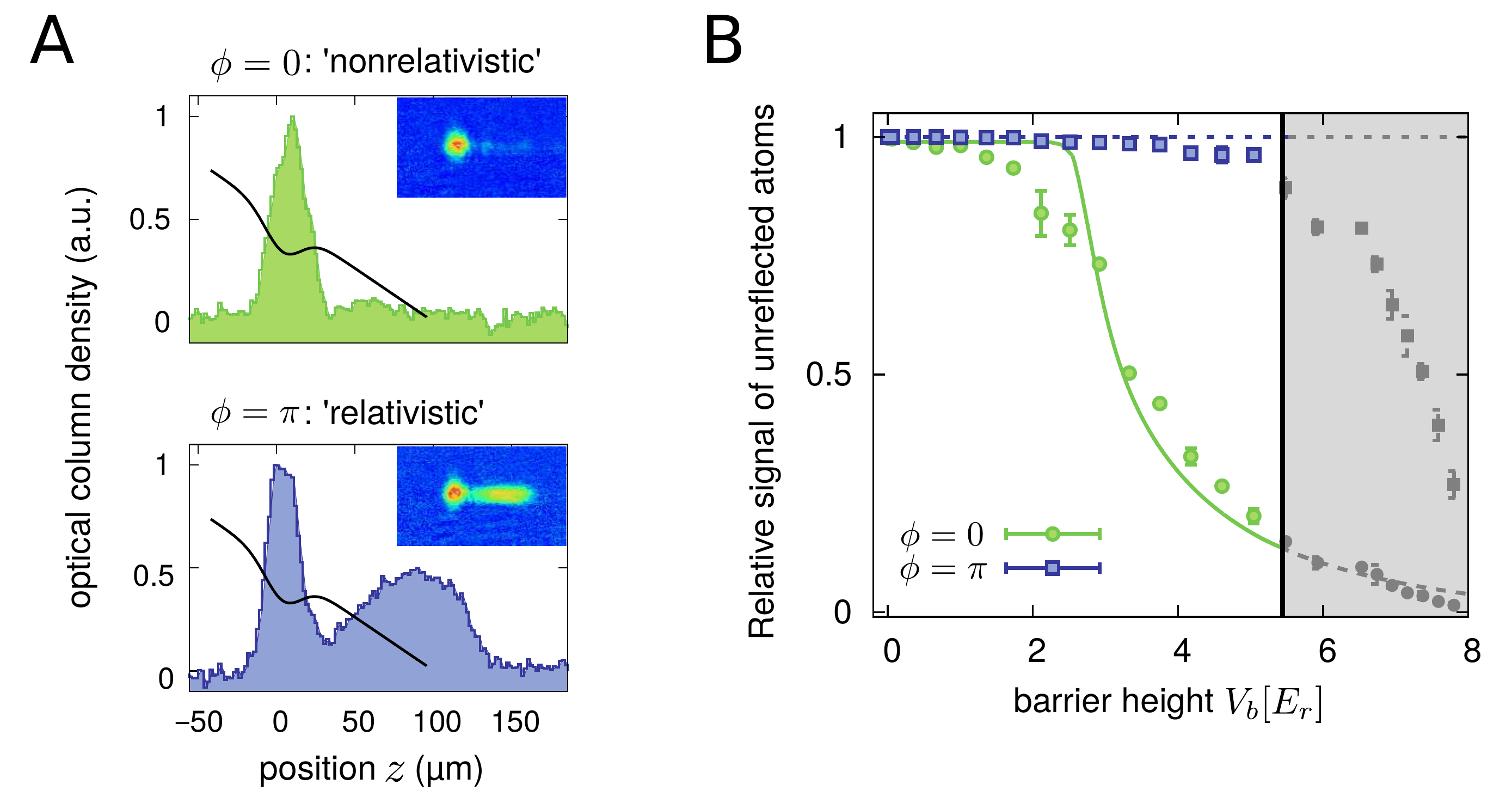}
  \caption{Experimental results for Klein tunneling of quasirelativistic atoms in the lattice. \textbf{A}: Linecuts and corresponding TOF images for a 'nonrelativistic' ($\phi=0$, green) and a 'relativistic' ($\phi=\pi$, blue) case. Only in the 'relativistic' case, atoms are able to overcome the potential barrier (indicated by black line). \textbf{B}: Relative signal of unreflected atoms versus the barrier height $V_b$. In the 'relativistic' case (blue squares), the signal of unreflected atoms remains constant within the range where the linear approximation of the energy dispersion relation remains valid (white background). In the 'nonrelativistic' case (green dots), the probability to overcome the potential barrier decreases with increasing potential height. The shown lines correspond to numerical simulations. The gray region indicates barrier height values for which the atomic velocity reaches the edge of the Brillouin zone of $\abs{q}=\hbar k$. Experimental parameters: $V_1=1.9 E_r$ and $V_2=0.2 E_r$.}
  \label{fig:4}
\end{figure}

Our experimental sequence begins by initially accelerating atoms after condensate preparation to an initial momentum of $2.9\hbar k$, which allows to load atoms into the second excited Bloch band of the lattice. This desired value of the momentum is achieved by initially leaving the atoms in free fall in the absence of a lattice potential until due to the earth gravitational acceleration the momentum has increased to $0.9\hbar k$, after which a Raman pulse imparts an additional $2\hbar k$ momentum increasing the atomic momentum to the desired value. Then, optical lattice and dipole trap potential are switched on and the atoms move towards the potential barrier. Fig. \ref{fig:4}A (top) shows experimental data for the atomic spatial distribution, as recorded by absorption imaging at a time $\SI{5}{ms}$ after preparation, for the case of a relative phase of $\phi=0$ between lattice harmonics, for which the splitting between the first and the second excited Bloch band is too large to allow for tunneling between bands, yielding an effectively nonrelativistic dispersion experienced by the atoms. Both the recorded absorption image (inset) and the solid green lines in the main plot of Fig \ref{fig:4}A (top), with the external potential indicated by the solid black line, show that almost all atoms remain in front of the potential barrier, as expected. On the other hand, for a relative phase of $\pi$ between lattice harmonics, our experimental data shows that atoms can be found behind the potential barrier, see Fig. \ref{fig:4}A (bottom), as attributed to the analog of Klein tunneling in the optical lattice system. If we choose the zero point of the energy scale to be at the crossing between the first and the second excited Bloch bands, see Figs. \ref{fig:1}B and \ref{fig:3} (middle plot), following the Feynman-St\"uckelberg interpretation of relativistic wave mechanics atomic population in the second excited band, i.e. above the crossing, correspond to particle-like excitation, in the below lying band of negative energy to a particle-like excitation of negative energy, being equivalent to a temporally forward propagating antiparticle excitation \cite{stueckelberg}. For the discussed case of a relative phase of $\pi$ between lattice harmonics, with an ultrarelativistic dispersion near the crossing, particles approaching the barrier loose kinetic energy on the rising edge and can reach to below the crossing of the first and the second excited band, i.e. below the Dirac point. They correspondingly can surpass higher potential barriers than in the 'nonrelativistic' case shown in Figs. \ref{fig:3} (top) and \ref{fig:4}A (top). This corresponds to the case of Klein tunneling, and our experiment simulates the conversion of a particle into a spatially backwards propagating antiparticle during the transmission of the barrier. On the tailing edge the band crossing is again passed, so that behind the barrier particle-like excitations are again observed. As applies to the case of Klein tunneling of electrons, this is equivalent to a double Landau-Zener tunneling between states of positive and negative energy respectively.

One prediction for Klein tunneling is that the barrier transmission should be independent of the barrier height, an issue in clear contrast to the case of nonrelativistic quantum tunneling. For our system, Fig. \ref{fig:4}B shows the measured relative signal of unreflected atoms, corresponding to the atoms undergoing Klein tunneling, versus the barrier height. The grey shaded region corresponds to barrier height values where atoms reach the edge of the Brillouin zone, i.e. only in the left white region the desired dispersion is reached, with increased accurancy near the bottom. If on the other hand the potential barrier is too shallow, atoms are able to leave the trap regardless of the energy dispersion relation. The blue dots are data for $\phi=\pi$, corresponding to a relativistic dispersion, for which the signal remains at a high, near constant value within the white region, illustrating the prediction of Klein tunneling not depending on the barrier height. In contrast, for $\phi=0$ (green dots), corresponding to the 'nonrelativistic' case, the signal decreases with increasing barrier height. Quantum tunneling here remains neglible due to the large width of the barrier, but the finite width of the kinetic atomic energy distribution here softens the otherwise expected sharp drop for $E<V_b$.

\section{Veselago lensing}

\begin{figure}
  \centering
  \includegraphics[width=0.9\textwidth]{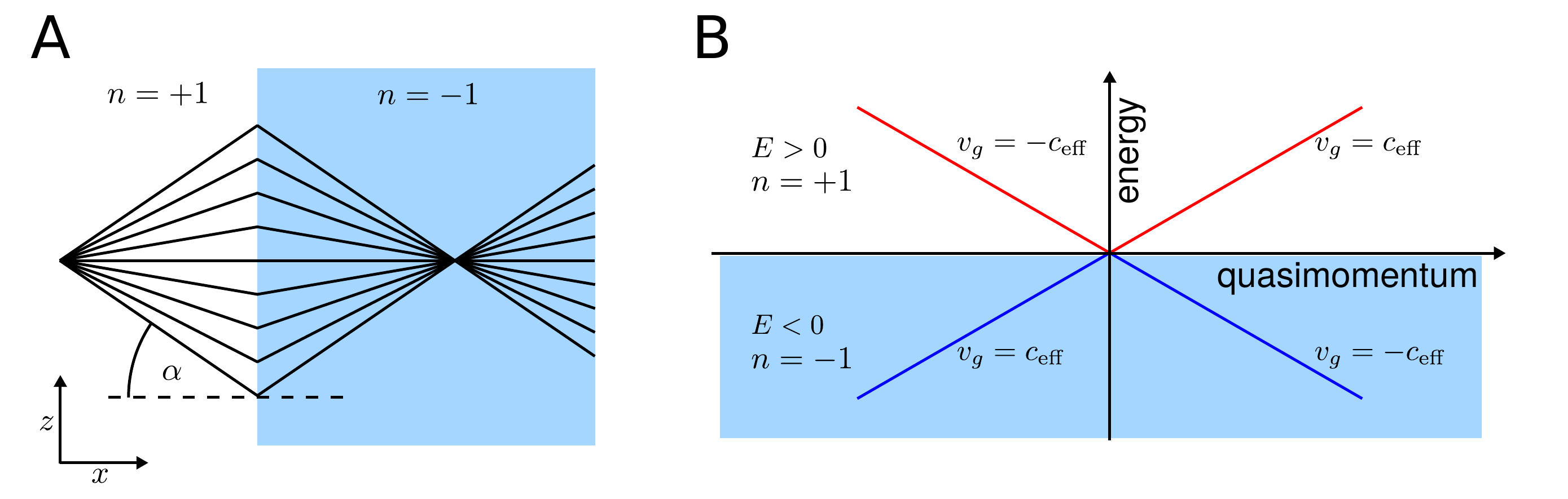}
  \caption{\textbf{A}: General principle of Veselago lensing. A pencil of diverging rays is converted to a pencil of converging rays by the refraction off a negative refractive index material. \textbf{B}: Relevant part of the dispersion relation for vanishing splitting between bands. States with positive energy here correspond to a positive index of refraction and vice versa. The experiment starts with atoms in the upper (drawn as red lines) branches of the dispersion. Negative refraction is realised by transferring atoms to states below the crossing (blue lines) using a four-photon Raman pulse.}
  \label{fig:5}
\end{figure}

Veselago lensing is  an imaging technique based on the refraction of rays during the transition from positive to negative refractive index materials. In this situation, negative refraction occurs. For electromagnetic waves a negative index of refraction can be realized e.g. in metamaterials, where the electric and magnetic properties can be taylored with below wavelength size structures. In negative index materials optical wave propagation is reversed, and a slab of such a material can focus a diverging pencil of rays into a spatially converging one, forming a Veselago lens \cite{Veselago}, see Fig. \ref{fig:5}A. This imaging technique strikingly differs from conventional, positive refractive index optics, where curved surfaces bend the rays to form an image of an object. For the latter, see also published work on the lensing of atomic matter waves with optical potentials imprinting a quadratic dispersion \cite{obertaler1,obertaler2}. Pendry has shown that the spatial resolution of a Veselago lens is not limited by the usual diffraction limit, because it can restore the evanescent waves \cite{pendry}. Veselago lensing has been proposed for matter waves, graphene material \cite{graphenveselago}, and dark state media \cite{darkstate} respectively, though no corresponding realizations have been reported. The essential ingredient of a Veselago lens is a relativistic (i.e. linear) dispersion relation, an issue that closely connects the here described work to the ones earlier discussed in this article.

In our experiment, Veselago lensing for matter waves is demonstrated using ultracold rubidium atoms in the bichromatic lattice potential with a linear dispersion relation near the crossing of the first and the second excited Bloch band \cite{leder14}. A Raman-$\pi$-pulse technique here transfers the atoms between different branches of the dispersion relation, and the relativistic lensing proceeds by a backwards propagation of atomic de Broglie waves on an energetically mirrored branch of the dispersion relation. In general, if we set the zero of the energy scale to the crossing, the energies of eigenstates in the first and second excited Bloch bands near the crossing is given by
\eq{eq:VeselagoEnergy}{E=\pm\sqrt{\rk{c_{\tn{eff}}\hbar k_z}^2 + \rk{\Delta E/2}},}
where $\hbar k_z=q$ denotes the quasimomentum of atoms along the lattice beam axis, and $c_{\tn{eff}}=2\hbar k/m$. In the following, we describe an experiment investigating the analog of Veselago lensing in a single dimension using atoms in the lattice, in the presence of an additional spatially slowly varying potential. For this experiment, we use a splitting of $\Delta E=0$ between the lowest two excited bands, which gives a linear dispersion relation near the crossing. We have also performed a ray-tracing simulation of a 2D Veselago lens where we realize different projections of a 2D Dirac cone by tuning of the energy gap $\Delta E$. For the latter, we refer to \cite{leder14}.

In a single dimension, two rays are possible, along and oppositely directed to the $z$-axis respectively. A schematic of the relevant part of the energy dispersion relation for $\Delta E=0$, as relevant for the 1D Veselago lensing experiment, is depicted in Fig. \ref{fig:5}B. We start the experiment by transferring atoms to the second excited band (red) with $E>0$ by using two-photon Raman pulses which transfer atoms from $\ket{m_f=0,\hbar k_z}$ to $\ket{m_f=-1,\pm 2 \hbar \kappa + \hbar k_z}$. By this, we realize two spatially diverging atomic paths of quasirelativistic velocity $\pm c_{\tn{eff}}$ opposed and along the optical lattice beam axis respectively. The group velocity $v_g=\partial E/ \partial q=\pm c_{\tn{eff}}$ is parallel to the corresponding wave vector along the z-axis. The situation resembles the propagation in a positive refractive index material, say $n=1$. After a propagation time $t_p=\SI{1.3}{ms}$, atoms are transferred to the lower excited energy band (indicated in blue in Fig. \ref{fig:5}B) with $E<0$ by using a short (broadband) four-photon Raman $\pi$-pulse which couples the states $\ket{m_f=-1, 2\hbar \kappa + \hbar k_z}$ and $\ket{m_f=-1, -2\hbar \kappa + \hbar k_z}$. Wave vector and group velocity are now opposed to each other and the atomic propagation is reversed, resembling a propagation in a negative refractive index material with $n=-1$. The paths meet again after a further propagation time, at which an image of the initial cloud is created. We superimpose an additional, spatially slowly varying confining potential (of $\SI{70}{Hz}$ trapping frequency) to the optical lattice potential.

\begin{figure}
  \centering
  \includegraphics[width=0.9\textwidth]{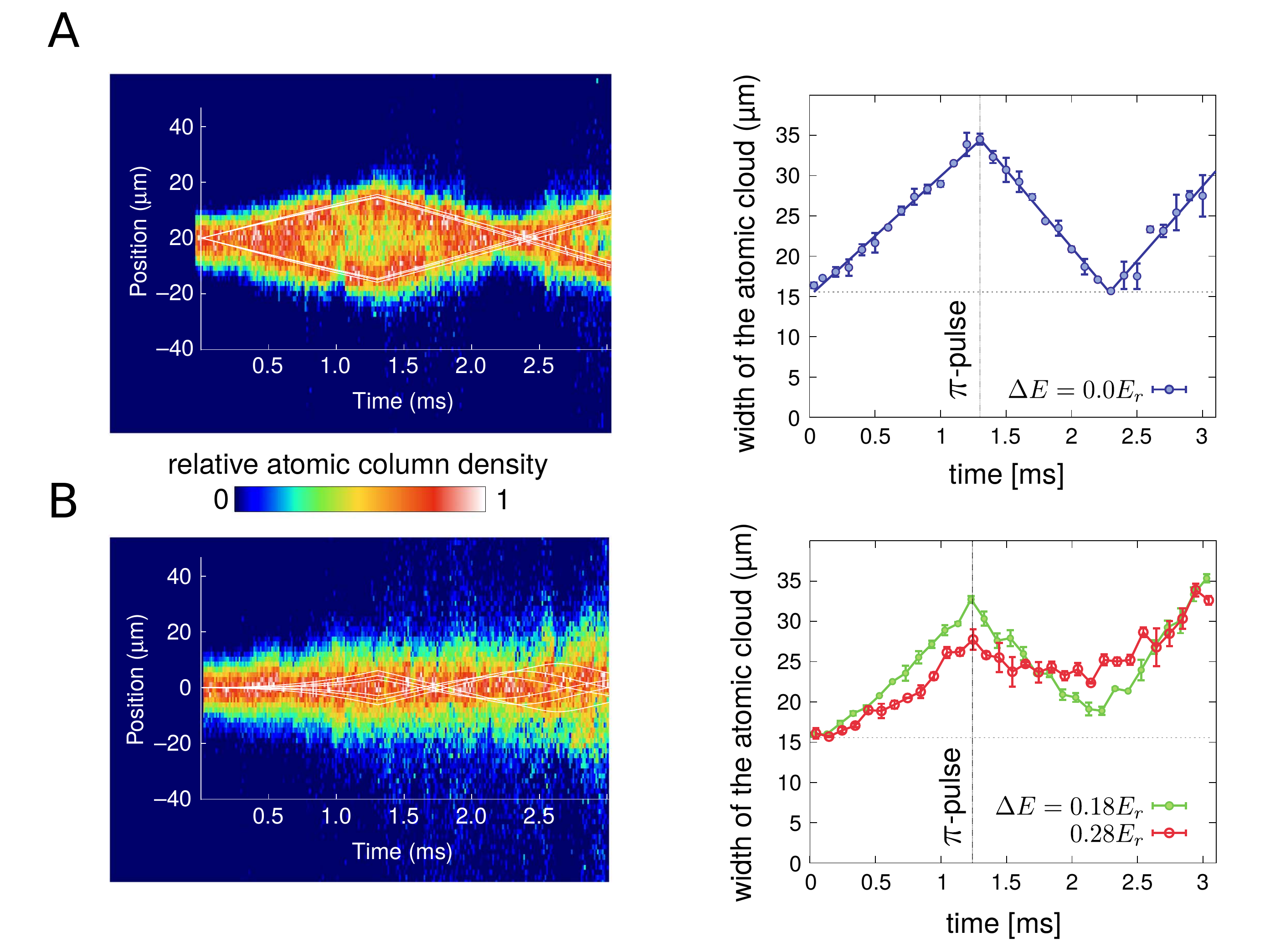}
  \caption{Experimental results for the 1D Veselago lens. The left panels show a series of absorption images, from which the atomic distribution can be extracted. The white lines show the results of a numerical simulation for different initial atomic momenta. The right panels show the width of the atomic cloud versus time.  \textbf{A}: For a linear dispersion relation, after approximately twice the expansion time before the four-photon Raman pulse is applied for refocussing, the width of the atoms equals the initial width within experimental uncertainties. \textbf{B}: For a nonvanishing splitting between the Bloch bands, the refocusing is less perfect. The observed cloud width near the second focus increases with a larger width of the splitting between bands $\Delta E$, see the graphs for the shown two values of the splitting $\Delta E$ at the right panel.}
  \label{fig:6}
\end{figure}

At the end of an experimental cycle, we detect the distribution of atoms in the $m_f=-1$ Zeeman state by a Zeeman state selective microwave transfer pulse followed by a resonant optical pulse recording an absorption image of the atoms transferred by the microwave pulse. Fig. \ref{fig:6}A shows a series of binned images (left) of such absorption imaging data for different propagation times in the lattice along with the corresponding width of the atomic cloud (right), as determined by Gaussian fits, for the case of a near linear dispersion near the band crossing. We observe that after $\SI{2.2}{ms}$ propagation time, the cloud width within experimental uncertainties is refocused to its initial width, as attributed to Veselago lensing. Sharp refocusing is here achieved even in the presence of an additional spatially varying potential, as understood from the (in an experimentally ideal situation) energy independence of the group velocity. The observed minimum width of the second focus is not exactly at twice the propagation time, as attributed to deviations from a perfectly linear dispersion. We note that in general one also expects an interference pattern of spatial periodicity $\lambda/4$ to occur, which is not observed experimentally due to the limited resolution of our imaging system.

For comparison, Fig. \ref{fig:6}B shows experimental data for a non-vanishing energy gap between bands, where the observed image is clearly broader than the initial atomic cloud size. Here the energy spread caused by the locally varying dipole potential causes different group velocities, as the group velocity for a finite splitting between bands is not energy independent anymore. The white lines in the left pictures of Fig \ref{fig:6}A,B are the result of numerical simulations for three different values of the wave vector $\hbar k_x$ correspondingly.

\section{Conclusion}
To conclude, ultracold atoms in variable optical lattice potentials can be used as a tool to observe relativistic wave equation predictions. We have considered the quantum simulation of such relativistic effects with ultracold rubidium atoms in a biharmonic optical lattice potential. The background of the presented experiments is that the dynamics of atoms in the biharmonic lattice near a crossing can be described by a one-dimensional Dirac equation. At the beginning of the experimental section of this article, we have presented a technique to diabatically load atoms into a specific Bloch state in the lattice near the crossing. We have then described two proof of principle experiments demonstrating the observation of relativistic wave equation predictions with ultracold atoms in the lattice. The analog of Klein tunneling of atoms, the penetration of a barrier without the exponential damping that is characteristic for nonrelativistic quantum tunneling, has been observed. Further, Veselago lensing for matter waves was demonstrated, an effect building upon negative refraction, where refocusing is achieved using a $\pi$-pulse technique transferring atoms into an energetically mirrored branch of the dispersion relation.

For the future, it is expected that cold atoms can be used to simulate nonlinear relativistic Dirac dynamics, as chiral confinement \cite{merkl} or predictions of interacting quantum field theory \cite{QFT}. Besides the use of spatially biharmonically modulated lattice potentials, as investigated in the present work, one may also use temporally modulated lattice potentials to engineer a desired quasirelativistic dispersion in the Floquet band structure of cold atoms \cite{grossert,struck}.

%%%%%%%%%%%%%%%%%%%%%%%%%%%%%%%%%%%%%%%%%%%%%

%%%%%%%%%%%%%%%%%%%%%%%%%%%%%%%%%%%%%%%%%%%%%%%

\end{document}